# Spatially homogeneous few-cycle compression of Yb lasers via all-solid-state free-space soliton management

**BINGBING ZHU[1†], ZONGYUAN FU[1†], YUDONG CHEN[1†], SAINAN PENG[1], CHENG JIN[2], GUANGYU FAN[3*], SHENG ZHANG[1], SHUNJIA WANG[1], HAO RU[1], CHUANSHAN TIAN[1], YIHUA WANG[1,4], HENRY KAPTEYN[5], MARGARET MURNANE[5], ZHENSHENG TAO[1*]**

[1] *Department of Physics and State Key Laboratory of Surface Physics, Fudan University, Shanghai 200433, China*
[2] *Department of Applied Physics, Nanjing University of Science and Technology, Nanjing, Jiangsu 210094, China*
[3] *Institut National de la Recherche Scientifique, Centre Énergie Matériaux et Télécommunications, Varennes, Quebec, Canada*
[4] *Shanghai Research Center for Quantum Sciences, Shanghai 201315, China*
[5] *Department of Physics and JILA, University of Colorado and NIST, Boulder, CO 80309, USA*
†*These authors contributed equally to this work.*
*\*Corresponding author: Guangyu.Fan@inrs.ca ; ZhenshengTao@fudan.edu.cn*

**Abstract:** The high power and variable repetition rate of Yb femtosecond lasers make them very attractive for ultrafast science. However, for capturing sub-200 fs dynamics, efficient, high-fidelity, and high-stability pulse compression techniques are essential. Spectral broadening using an all-solid-state free-space geometry is particularly attractive, as it is simple, robust, and low-cost. However, spatial and temporal losses caused by spatio-spectral inhomogeneities have been a major challenge to date, due to coupled space-time dynamics associated with unguided nonlinear propagation. In this work, we use all-solid-state free-space compressors to demonstrate compression of 170 fs pulses at a wavelength of 1030 nm from a Yb:KGW laser to ~9.2 fs, with a highly spatially homogeneous mode. This is achieved by ensuring that the nonlinear beam propagation in periodic layered Kerr media occurs in soliton modes and confining the nonlinear phase through each material layer to less than 1.0 rad. A remarkable spatio-spectral homogeneity of ~0.87 can be realized, which yields a high efficiency of >50% for few-cycle compression. The universality of the method is demonstrated by implementing high-quality pulse compression under a wide range of laser conditions. The high spatiotemporal quality and the exceptional stability of the compressed pulses are further verified by high-harmonic generation. This work represents the highest efficiency and the best spatio-spectral quality ever achieved by an all-solid-state free-space pulse compressor for few-cycle-pulse generation.

## 1. Introduction

The generation of laser pulses of ~1-10 cycles in duration has enabled researchers to explore the fastest electron dynamics in atoms [1–3], molecules [4–6] and solids [7–10]. Moreover, intense femtosecond laser pulses have opened up new opportunities for generating attosecond XUV pulses via high-harmonic generation (HHG) [11–13] as well as for accelerating electron bunches to relativistic energies [14]. Experimentally, few-cycle optical pulses can be generated either directly from a laser, or via post-compression of femtosecond laser pulses [15] or more recently via optical parametric chirped-pulse amplification [16]. However, achieving high-quality spatial and temporal profiles and stable pulse compression over a large range of pulse energies can be challenging.

One key feature of many pulse-compression techniques is the interplay between supercontinuum generation (SCG), which is enabled by self-phase modulation (SPM), and compensation of negative group-delay dispersion (GDD) [17,18] For high-quality and high-efficiency few-cycle pulse generation, it is essential to manage the interaction of intense light with the nonlinear medium to achieve a stable ultrabroad spectrum, with high spatio-spectral-temporal quality. While this can be achieved at lower pulse energies by guiding a self-phase-modulated beam in a fiber and distributing the nonlinear phase over an extended propagation length [19–21], it is extremely challenging to achieve the same high-quality output at higher pulse energies via pulse compression in solids, where free-space propagation is required. However, this latter scheme possesses several unique advantages in practice compared to gas media. First, it does not require complex and costly vacuum systems and gas-vacuum interfaces. Second, it is insensitive to pointing fluctuations, which leads to more stable and robust performance. Third, the free-space geometry makes it flexible and easy to adjust for different laser conditions. The major drawback associated with this method is the nonlinear spatial inhomogeneities that are induced, which can result in ~50-60% loss of output power [22–24], in addition to undesirable effects caused by a complex spatial mode and spatial chirp.

In recent years, advances in diode-pumped Yb-doped femtosecond lasers [25–27] have enabled high power scaling with adjustable pulse energy and repetition-rate. However, the intrinsic narrow bandwidth of the Yb-doped gain medium limits the pulse duration from these lasers to >150 fs, precluding their application for capturing very fast processes. Moreover, compared to Ti:Sapphire lasers, it is even more challenging to compress Yb lasers to the few-cycle level, because of the larger compression ratio and other undesirable effects when the pulse duration is long (e.g. lower damage threshold of materials and more plasma accumulation during light-matter interaction, etc). To date, many post-compression schemes have been implemented to Yb lasers, mostly using gas media [28–38], and a few using solids [24,39–42]. Although few-cycle pulse duration (<10 fs) has been demonstrated with solid-state pulse compression using a free-space geometry [24,42,43], the usable optical energy is low (typically ~20% of the input power) due to spatially inhomogeneous nonlinear effects [22–24] and the need to implement multiple compression stages. Furthermore, to date implementing such solid-state free-space compressors is highly empirical with numerous free parameters. For instance, although nearly single-cycle compression of a Yb laser was demonstrated in Ref. [42] with two cascading compressors consisting of few thin material plates, the similar compression scheme can only deliver either a much longer duration [39] or rather low efficiency [43] in other works, even when the laser conditions are similar.

In a recent work, it was shown that the space-time-coupled propagation of intense Yb-laser pulses in periodic layered Kerr media (PLKM) can occur in discrete spatial soliton modes, when diffraction and nonlinear self-focusing are well balanced. This can provide a guiding mechanism for the propagation of self-phase-modulated pulses in an all-solid-state free-space apparatus [44]. Because soliton formation is wavelength dependent, one might anticipate that it would be extremely challenging to extend such a soliton-mode approach to SCG and pulse compression of few-cycle pulses. In this work, we develop a predictive method for implementing all-solid-state free-space PLKM compressors and show that it can be extended to the ultra-broadband few-cycle regime. High-efficiency and high-quality compression to few-cycle pulses (~9.2 fs) is demonstrated utilizing 1030 nm 170 fs pulses from a Yb:KGW laser and a two-stage PLKM compressor. Remarkably, our experimental and theoretical results show that the spectral broadening is spatially uniform, which is in contrast to previous results [22,23,37,42,45]. We achieve this by implementing nonlinear beam propagation in PLKM in soliton modes, as well as by limiting the nonlinear phase through each material layer to <1.0 rad. To our knowledge, our results represent the highest efficiency (>50%) and spatio-spectral homogeneity (~0.87) ever generated by all-solid-state free-space few-cycle pulse compressors. The universality of the method is demonstrated by successful compression of laser pulses with energies, repetition rates and pulse durations ranging by more than one order of magnitude (energy and repetition rate from 80 µJ, 100 kHz to 1 mJ, 10 kHz; pulse duration from 170 fs to 9 fs) at a 10-W level. Finally, to demonstrate the high spatio-temporal quality of the few-cycle compressed pulses, we drive the extreme nonlinear process of HHG, and observe an extension of the cut-off photon energy together with exceptional stability of the XUV radiation, which is facilitated by the high-quality all-solid-state pulse compression.

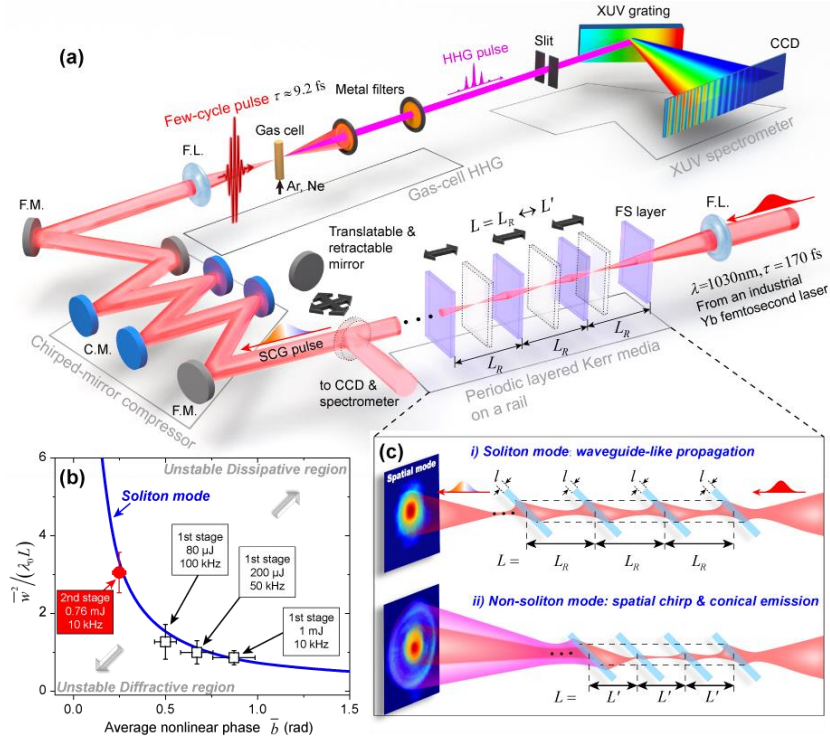

Fig. 1. (a) Schematic of the experimental setup. The long femtosecond pulse is focused into PLKM for SCG. A translatable and retractable mirror is used to deflect the beam for monitoring the spectra and beam profiles in the far field, from which the appropriate condition $L=L_R$ can be determined. The HHG is generated by focusing the compressed pulses into a gas cell. F.L.: focusing lens; F.M. folding mirror; C.M.: chirped mirror. (b) The universal relationship between the normalized beam radius squared, i.e., $\bar{w}^2/(\lambda_0 L)$, and the average nonlinear phase $\bar{b}$ for the soliton modes predicted by the FKD-FLI model [44]. The symbols are the experimental results under different conditions. (c) The illustrations of the beam propagation under i) the soliton mode and ii) the non-soliton mode under different $L$. When $L$ is short, conical emission and spatial chirp can be clearly observed.

## 2. Results and discussion

### 2.1 Complete recipe for constructing a PLKM pulse compressor

Figure 1(a) shows a schematic of our experimental setup. Pulse compression is realized by an all-solid-state pulse compressor consisting of PLKM and chirped-mirror compressors. The HHG radiation is then generated by focusing the compressed pulses into a gas-cell, and the HHG spectrum is characterized with an XUV spectrometer. (See Supplementary Section S1 for the detailed experimental setup.) In our experiments, the input laser is a Yb:KGW laser (PHAROS, Light-Conversion) with a central wavelength of $\lambda_0$=1030 nm, full-width-at-half-maximum (FWHM) pulse duration of $\tau$=170 fs, and average power is 10 W. The repetition rate can be adjusted between 200 kHz and 10 kHz, with the corresponding pulse energy varying from 50 µJ to 1 mJ.

In our experiments, we focused the laser beam (p polarization) into the PLKM, as shown in Fig. 1. The PLKM is composed of fused silica (FS) layers with equal thickness $l$ and spacing (PLKM period) $L$. The layers are mounted at Brewster's angle to minimize the transmission loss. As shown in Fig. 1(a), their positions can be individually adjusted along the beam direction using a rail with an accuracy of <1 mm. The key idea is to manage the nonlinear beam propagation in the PLKM by adjusting $L$ to form the soliton modes [Fig. 1(c)] by using the following procedure:

1) We determine the focusing condition. The intensity at the beam focus should be low enough to avoid strong ionization or filamentation in air. For the wide range of pulse energies and repetition rates used in our experiments, we simply focus the beam in air to a $1/e^2$ waist of 120 - 140 µm using a 1m focusing lens, which corresponds to a peak intensity of at most ~20 TW/cm² (for a pulse energy of 1 mJ) at the beam focus calculated at normal incidence.

2) The first layer of the PLKM is then placed at or behind the beam focus, where the intensity is below the damage threshold of the Kerr medium (~3.4 TW/cm² for FS at $\tau$=170 fs).

3) The layer thickness, $l$, is then selected to yield a nonlinear phase from the first layer ($b_0$) between 0.8 and 1.5 rad. The nonlinear phase is given by $b_0 = \frac{2\pi}{\lambda_0} n_2 \alpha l I_0$, where $n_2$ is the nonlinear refractive index of the Kerr medium, $w_0$ the beam radius, $I_0$ the intensity, and $\alpha$ is the intensity correction factor ($\alpha$≈0.83 for FS, see Supplementary Section S1). For the FS layers used in our experiments, $l$ ranging from 100 to 800 µm were selected.

4) The rest of the layers are then placed at equal spacings $L$, with the initial value estimated from $L_i = \frac{b_0 w_0^2}{2.0 \lambda_0}$. We note that $L_i$ is intentionally and empirically chosen to be smaller than the condition for the soliton modes: $L = \frac{\bar{b} \bar{w}^2}{0.8 \lambda_0}$ [44], where $\bar{b}$ and $\bar{w}$ stand for the average nonlinear phase and average beam radius on all the layers, respectively.

5) The formation of the soliton modes in the PLKM is then achieved by adjusting $L$, while the variations of the far-field beam size and the axial spectrum are simultaneously monitored (see below). It is important to note that the camera for measuring the beam size must be kept the same distance downstream from the last layer as $L$ varies. This is achieved in our experiments by a translatable and retractable mirror installed on the rail, as illustrated in Fig. 1(a).

6) It is important that ~10 layers are typically used to keep the average nonlinear phase sufficiently low (see below). The number of layers is then refined by removing the ones at the end of the setup that do not contribute to the spectral broadening.

We find that the output beam can exhibit a nice transverse spatial mode when we choose $b_0$≤1.5 rad, while complex spatial chirp starts to appear and affects the beam focus when $b_0$>1.5 rad. In contrast to past work on pulse compression that used only few thin Kerr plates [22,39,42,43,45,46], we construct a periodic structure with ~10 material layers with the equal thickness and separation to form a resonator-like geometry. This not only facilitates the formation of spatial solitons, but also significantly reduces the number of free parameters when constructing the apparatus.

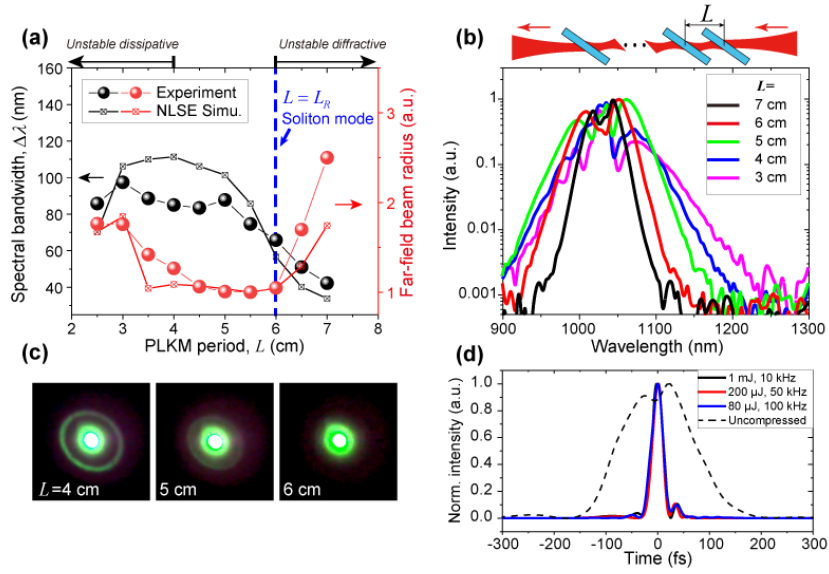

Fig. 2. (a) The spectral bandwidth and the normalized far-field beam radius as a function of the PLKM period $L$ obtained from our experiments and NLSE simulations (simu.). The PLKM is composed of 13 FS layers with $l = 800$ μm. The input pulse energy is 200 μJ at 50 kHz. The soliton mode can be identified and labelled by the blue dashed line. (b) The experimentally measured on-axial spectra of the output beam under different $L$ corresponding to (a). (c) The far-field spatial mode of the output beam measured by a laser viewing card (Thorlabs, VRC4) and the images are taken by a commercial camera. (d) The temporal intensity profiles of the compressed pulses under different laser conditions (1mJ, 10 kHz; 200 μJ, 50 kHz; 80 μJ, 100 kHz). The dashed line represents the uncompressed pulse from the laser.

To demonstrate the flexibility of our method, high-quality pulse compression was implemented under a wide range of laser conditions: 1 mJ, 10 kHz; 200 μJ, 50 kHz; and 80 μJ, 100 kHz. Taking the input pulse energy of 200 μJ at 50 kHz as an example, we choose the FS layer thickness to be $l$=800 μm. The first layer is positioned ~10 cm behind the beam focus. Under these conditions, the nonlinear phase on the first layer, $b_0$, is estimated to be ~1.32 rad and the initial spacing $L_i$ is thus ~4 cm. Following Step 5 above, we measure the variations of the far-field beam size and the spectral bandwidth ($\Delta\lambda$) of the output beam under different $L$ after passing through 13 FS layers [Fig. 2(a)]. Here, $\Delta\lambda$ is defined as the spectral width containing 75.8% of the total optical energy. The representative spectra are summarized in Fig. 2(b). Notably, as $L$ reduces from 7 cm to 2.5 cm, we observe a non-monotonic change of the far-field beam radius, which can be well reproduced by our nonlinear-Schrodinger-equation (NLSE) simulations conducted under the same experimental conditions [Fig. 2(a)]. In the NLSE simulations, the dispersive, the nonlinear Kerr and Raman effects are considered [44,47] (see Supplementary Section S2). The corresponding space-time-coupled beam propagation under different $L$ is illustrated in Fig. 3(a)-(c).

According to our simulations, the results in Fig. 2(a) can be understood by considering the interplay between diffraction and self-focusing effects. On the long-$L$ side, the intensity on the second layer is already weak, and the nonlinear self-focusing is thus not sufficient to confine the transverse beam expansion on the consecutive layers, leading to a diverging beam size in the far field, as shown in Fig. 3(c). This corresponds to the "*Unstable diffractive region*" when $L$>6 cm [44] [Fig. 2(a)]. On the other hand, when $L$ is short to the condition that the focal point of self-focusing may lie on or close to one of the consecutive plates. Although the repetitive strong Kerr lensing can still provide confinement on the transverse size of the central mode, strong conical emission and spatial chirp can be observed [Fig. 3(b)], resulting in inevitable spatial and temporal losses [44]. This corresponds to the "*Unstable dissipative region*" when

$L$<4 cm [Fig. 2(a)]. According to our NLSE simulations, the beam diffraction and nonlinear self-focusing can well balance each other when $L$= 6 cm, which allows the formation of the discrete spatial soliton with a stable beam size on the 13 FS layers of PLKM in this specific case [Fig. 3(a)].

We note that $\Delta\lambda$ can be larger when 4 cm<$L$<6 cm (the "*quasi-stable region*" [44]), e.g. at $L$=5 cm, $\Delta\lambda$ is broader than that at $L$=6 cm as shown in Fig. 2(a) and (b), which presumably can be compressed to a shorter pulse. However, the spatial beam quality for $L$=5 cm has already deteriorated due to the spatial chirp and conical emission, which is manifested by the appearance of optical rings around the central mode [Fig. 2(c)]. Moreover, the complex space-time coupling leads to higher-order dispersion [44,48], which would require custom-designed chirped mirrors or pulse shapers to compress [42,45]. As a result, we choose $L$=6 cm ($L_R$) for the final setup. Experimentally, this corresponds to the $L$ value for the minimum beam radius, approaching from the long-$L$ side [Fig. 2(a)]. Under this condition, $\Delta\lambda$ of ~70 nm along with guided beam propagation in the PLKM can be simultaneously realized, leading to SCG with high efficiency and high spatio-spectral quality in a free-space setup.

Following the above procedure, the soliton modes can be routinely generated for other laser conditions (1 mJ, 10 kHz and 80 μJ, 100 kHz), where the corresponding key parameters are listed in Table 1. Here, $\bar{b}$ is estimated by the spectral-broadening ratio and the number of implemented layers [49] (see Supplementary Section S3), and its value is generally smaller than $b_0$ (e.g. $\bar{b}$ ~0.67 rad for the above case of 200 μJ, 50 kHz). This could be attributed to the beam energy loss during the propagation, as well as the self-adjustment of spatial mode in the first few layers. The optimum PLKM conditions for the different laser parameters are also summarized in Fig. 1b. Remarkably, they all exhibit excellent agreement with the soliton modes theoretically predicted by the Fresnel-Kirchhoff diffraction model and the Fox-Li iteration (FKD-FLI, see Supplementary Section S3) [44], which further corroborates that our approach can reliably generate the soliton modes under different laser conditions.

**Table 1: List of the PLKM parameters for different laser conditions**

| Input pulse energy (μJ) | Repetition rate (kHz) | $l$ (μm) | $L_R$ (cm) | Beam size (on the 1st layer) (μm) | Average nonlinear phase, $\bar{b}$ (rad) | Number of layers |
|---|---|---|---|---|---|---|
| **1000** | **10** | 400 | 13.0 | 340 | 0.87 | 10 |
| **200** | **50** | 800 | 6.0 | 250 | 0.67 | 13 |
| **80** | **100** | 800 | 2.5 | 130 | 0.50 | 16 |
| **760**[a] | **10**[a] | 100 | 9.5 | 545 | 0.25 | 10 |

[a] The conditions of the second-stage compressor.

The spectrally broadened SCG pulses are then compressed using a set of chirped mirrors (Ultrafast Innovations, PC1611) which supplies an appropriate amount of negative GDD [Fig. 1(a)]. We note that, because no spatial filtering is necessary in our method, the overall efficiency of the compressor can reach ~85%, which includes ~10% loss from the reflections on the Kerr media and ~5% loss on the collimation optics and chirped mirrors. The temporal profiles of the compressed pulses from the single-stage PLKM compressor are shown in Fig. 2(d), which are measured by second-harmonic-generation frequency-resolved optical gating (SHG-FROG) [50]. The compression to a FWHM duration of $\tau$≈22 fs can be routinely and reliably achieved for a wide range of laser conditions. The $M^2$ measurements yield typically $M_x^2$ =1.56 and $M_y^2$ =1.64 for the compressed beams (see Supplementary Section Fig. S3).

*2.2 Rules for constructing a spatially homogeneous free-space compressor*

For all-solid-state free-space SCG and pulse compression, spatio-spectral homogeneity has been a major challenge [21–24]. Because the SPM process is intensity dependent, one might anticipate that the spectral broadening of a Gaussian beam should be inhomogeneous in space:

while broad spectra can be generated in the beam center where the irradiance is high, the spectra at the beam wings with much lower intensity will hardly be changed. Indeed, it has been shown that the inhomogeneity of spatial nonlinearity can lead to 50-60% power loss of the output power when the laser beam is sent through a single bulk material [23,24] or through few thin material plates [22].

To evaluate the spatio-spectral homogeneity of the output beam in our experiments, we measure the spectral profiles along the radial coordinate ($r$) by selecting a small portion of the beam in the far field with a round aperture [inset of Fig. 3(d)]. The PLKM period is fixed at the optimum condition $L=L_R$. The spectral overlap integral $V(r)$ is given by [41]:

$$V(r) = \frac{\left\{\int [I(\lambda,r)I(\lambda,0)]^{\frac{1}{2}} d\lambda\right\}^2}{\int I(\lambda,r)d\lambda \cdot \int I(\lambda,0)d\lambda}, \quad (1)$$

where $I(\lambda,r)$ is the spectral intensity at the radial coordinate $r$. The average overlap integral across the beam is then evaluated by $\bar{V} = [\int V(r)I(r)rdr]/[\int I(r)rdr]$. In Fig. 3(d), we plot the experimentally measured $V(r)$ along with the optical intensity. Remarkably, within the far-field $1/e^2$ beam radius ($r_0$) the spectral overlap can maintain at >0.9, and the integration over the range of $2r_0$ yields $\bar{V}\approx0.95$, indicating a spatially uniform spectral broadening across the beam.

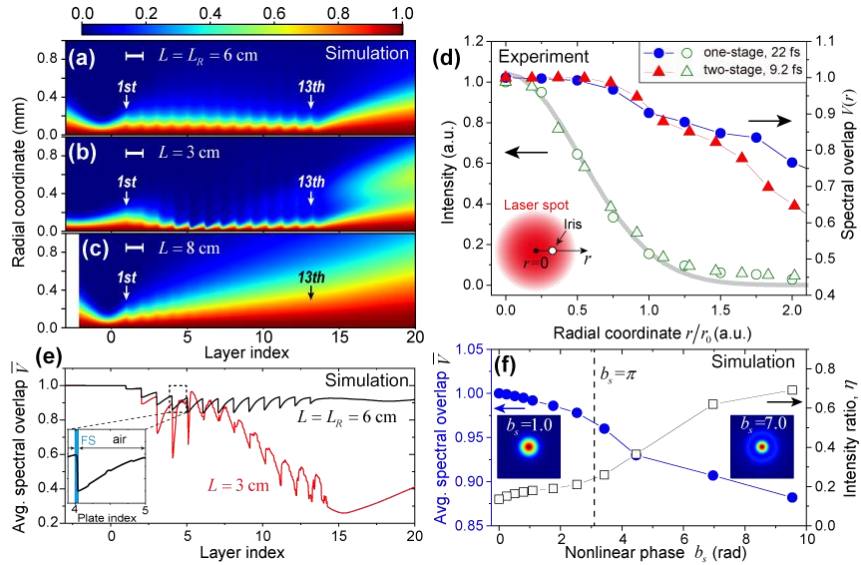

Fig. 3. (a)-(c) The variation of the beam radial intensity as the beam propagates through the PLKM with $L$=6 cm ($L_R$), 3 cm and 8 cm, respectively, obtained from the NLSE simulations. (d) The beam intensity and spectral overlap $V(r)$ along the radial coordinate $r$ for the compressed pulses under the optimum conditions of the one-stage (~22 fs) and two-stage (~9.2 fs) compressors. Inset: illustration of the spatial-filtering experimental setup. (e) The evolution of the average (avg.) spectral overlap integral, $\bar{V}$, obtained from the NLSE simulations. The simulation conditions are the same as (a) for $L$=6 cm and (b) for $L$=3 cm. Inset: the zoom-in view of the $\bar{V}$ variation within and after a single FS plate. (f) The NLSE simulations of $\bar{V}$ and the intensity ratio $\eta$ after the laser beam passes through a single FS plate with $l$ = 800 μm. Inset: the spatial modes of the output beam with $b_s$=1.0 and 7.0 rad.

To compare our method with previous results [22–24,42,45], in Fig. 3(e), we plot the evolution of $\bar{V}$ under different $L$ values obtained from our NLSE simulations. Under the soliton mode ($L=L_R$=6 cm), although dramatic decreases of $\bar{V}$ can be observed within each FS plate,

$\bar{V}$ is then greatly recovered through the free-space propagation in air afterwards until the next material layer [inset of Fig. 3(e)]. By placing the consecutive layer at the appropriate position and repeating this process under the soliton mode, $\bar{V}$ can be maintained at ~0.92 after propagating through 13 thin FS layers, which is in excellent agreement with our experimental observation. In stark contrast, $\bar{V}$ is reduced to less than 0.4 when $L$=3 cm and the beam propagation deviates from the soliton mode. This is consistent with the complex spatial mode observed in both the experiments [Fig. 2(c)] and the simulations [Fig. 3(b)].

The dramatic drop of $\bar{V}$ inside the FS plates [inset of Fig. 3(e)] is a result of the spatial inhomogeneity of the nonlinear effect. It is obvious that, when the nonlinear effect is too strong, $\bar{V}$ will not recover after the single layer [23]. In Fig. 3(f), we simulate the SPM effect on the beam spatial homogeneity through a single FS layer. The thickness ($l$=800 μm) and the focusing geometry is kept the same as for the previous results. Here, the nonlinear phase on a single FS layer $b_s$ is calculated by considering the average laser intensity inside the layer, and the intensity ratio $\eta$ is defined as the ratio of the optical power for $r$>$r_0$: $\eta = [\int_{r_0}^{\infty} I(r)rdr]/[\int_0^{\infty} I(r)rdr]$. The free-space propagation after the layer is long enough to yield stable $\bar{V}$ and $\eta$ under each intensity. Our results show that the reduction of $\bar{V}$ accelerates when $b_s > \pi$, which is accompanied by a rapid increase of $\eta$, indicating the appearance of the non-Gaussian spatial mode [inset of Fig. 3(f)]. This complex spatio-spectral behavior can be intuitively understood by considering the destructive interference between the radiations from the beam center and the wings when $b_s$ approaches $\pi$. Previously, with the coupled-mode theory, Milosevic *et al.* pointed out that the accumulated nonlinear phase in a single pass through a bulk medium should be confined to the order of unity, in order to preserve the high spatio-spectral homogeneity [21]. Our results are consistent with this conclusion, and the conditions realized by our recipe all yield $\bar{b}$ <1.0 rad, which is the key for the high spatio-spectral quality.

More importantly, this result also reveals several important rules for constructing such a free-space compressor and reaching high spatial-spectral quality. First, it defines the rule on the minimum number of material layers. For example, for pulse compression from 170 fs to 22 fs with a compression ratio of ~7.7, it requires a total nonlinear phase of ~8.7 rad [49]. If we limit the nonlinear phase on each material layer to be <1.0 rad, at least 9 layers are required to preserve the high spatio-spectral quality of SCG. Compression with fewer layers would inevitably lead to temporal and spatial losses, which also explains the conical emission and spatial chirp commonly observed in previous studies with a bulk layer or few thin plates [22–24,42,45]. We note that this rule cannot be relaxed by simply changing the layer material, thickness, or the experimental geometry, and there is no easy way to recover the deteriorated beam quality in practice. Second, the positions of the layers are important, since the nonlinear phase can easily go beyond the threshold of 1.0 rad, when the positions are inappropriate. For example, for $L$=3, the nonlinear phase exceeds 2.0 rad after the 4[th] layer, as illustrated in Fig. 3(b). With our recipe, by limiting the initial nonlinear phase $b_0$~1.0 rad and managing the beam propagation into the soliton mode to distribute the nonlinear phase as evenly as possible, the output beam with broad and spatially uniform spectrum can be routinely realized.

*2.3 Generation of high-quality few-cycle pulses*

A second-stage PLKM is constructed to further compress the pulses to few cycles. Here, we employ the compressed pulses from the first-stage compressor (~22 fs, ~0.76 mJ, 10 kHz) as the input. The beam is focused into the second-stage PLKM with a beam waist of ~500 μm. The layer thickness is selected to be $l$=100 μm, which yields $b_0$ ~ 1.0 rad. The same procedure in Section 2.1 is applied, which yields $L_R$≈9.5 cm for the balanced propagation (see Supplementary Section S5). The final conditions of the second-stage compressor are summarized in Table 1 and Fig. 1(b). Remarkably, the resultant condition also exhibits excellent agreement with the FKD-FLI model even when the incident pulse is broadband. The ultra-broadened spectrum after the second-stage PLKM is plotted as the blue line in Fig. 4(a). The spectrally broadened pulses are compressed by a chirped-mirror set and a wedge pair (see Supplementary Fig. S1), and a FWHM pulse duration of 9.2 fs can be achieved, as shown in Fig. 4(b). The experimental and reconstructed SHG-FROG traces are shown in Fig. 4(c) and (d), respectively, and the spectral marginal is also plotted in Fig. 4(a).

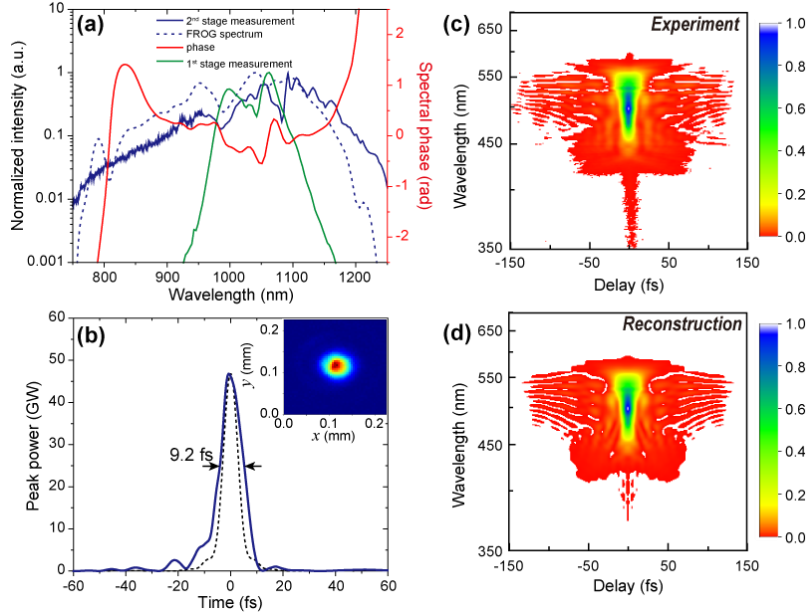

Fig. 4. (a) The measured, FROG-reconstructed spectra and the retrieved phase of the few-cycle pulses from the second-stage compressor. The spectrum from the first-stage compressor is shown as the green line for comparison. (b) The reconstructed intensity profile of the few-cycle pulses after the two-stage compressor with $L$=9.5 cm. The transform-limited pulse profile is plotted as the dashed line. Inset: the output beam profile of the few-cycle pulses after the focus. (c) and (d)The measured and reconstructed FROG traces for the few-cycle pulses.

The overall power transmission of the second-stage compressor is ~80%, corresponding to an output pulse energy of ~0.6 mJ. By constructing the second-stage compressor with the method, we do not observe excessive conical emission and the beam can be well focused to a small spot size [inset of Fig. 4(b)]. Yet, through more detailed spatio-spectral analysis (see Section 2.2), we find that some spatial chirp still appears, as shown in Fig. 3(d), which makes $V(r)$ degrade faster at large $r$ compared to the first-stage results. Still, the value of $V(r)$ can be preserved to >0.9 within the $1/e^2$ beam radius, and $\bar{V}$ across the entire beam yields ~0.87. Thus, the compression efficiency of >50% and a peak power of ~50 GW can be achieved for the spatially uniform few-cycle pulses [Fig. 4(b)]. We note that, to the best of our knowledge, our results represent the highest efficiency and the best spatio-spectral quality of few-cycle pulses ever achieved in an all-solid-state free-space compressor.

The residual spatial chirp could be a result of the complex interplay between the dispersion and the spatio-temporal coupling associated with the ultra-broadened spectrum [51]. Unlike the propagation of the initially narrow-band pulse of $\tau$=170 fs, for which the temporal profile is mostly stable throughout the propagation under the soliton mode [44], the initially broad-band pulse of $\tau$≈22 fs can be strongly chirped due to the dispersion accumulated from both the Kerr media and the SPM process, not to mention that the spectrum is further increased to span a nearly octave spectral range in the second-stage PLKM. As a result, light with different wavelengths may experience a slightly different strength of the Kerr nonlinearity, resulting in an inhomogeneous spatio-spectral distribution of the output beam. However, our results, as well as their agreement with the FKD-FLI model [Fig. 1(b)], interestingly, indicate that the optimum condition determined by the method of Section 2.1 can still help to optimize the space-time-coupled beam propagation and minimize the loss caused by the spatial inhomogeneity, even in the ultra-broadband few-cycle regime.

*2.4 High-harmonic generation with compressed pulses*

The quality of the compressed few-cycle pulse can be rigorously evaluated by the HHG process, which is an extreme nonlinear-optical process that is very sensitive to the spatio-temporal properties of the driving pulses. The compressed pulses from the single-stage ($\tau\approx$22 fs) and the two-stage ($\tau\approx$9.2 fs) compressors are focused onto a gas target inside a vacuum chamber. A simple cylindrical gas cell with an inner diameter of 1.5 mm supplies argon and neon as the gas targets. The gas pressure is optimized for the phase-matching conditions, which yields ~50 and 120 Torr in argon for $\tau\approx$9.2 fs and 22 fs, respectively. The phase-matching pressures in neon are ~150 and 520 Torr for the two pulse durations. The beam waist at the focus is 50-60 μm. The harmonic spectrum is characterized with an XUV spectrometer after blocking the driving optical beams with 300 nm aluminum and/or 200 nm zirconium filters [Fig. 1(a)]. In Fig. 5(a), we plot the HHG spectra driven by the compressed pulses: 0.76 mJ, 22 fs and 0.25 mJ, 9.2 fs with a 10 kHz repetition rate, which yields a similar intensity at the focus (~400 TW cm$^{-2}$). The cut-off photon energy of the XUV radiation in both argon and neon is greatly extended when the driving pulse duration is reduced to few cycles, which can be attributed to the suppression of the gas ionization [30] and the nonadiabatic HHG process [52,53]. This proves the high spatiotemporal quality of the compressed few-cycle pulses. The HHG spectra in Argon (Fig. 5(a), 1 μm, 0.76 mJ, 22 fs) compares well with Ti:sapphire-driven OPAs (1.3 μm, 0.9 mJ, 700 torr, 30 fs), which yields a cutoff photon energy of 110 eV, and a photon flux approaching $10^{10}$ photons/s in 1% bandwidth at a 1 kHz repetition rate [54]. For Neon, the single-atom and phase matched photon energy cutoffs are slightly lower (130eV compared with ~190eV), which is likely due to ionization-induced defocusing and lower gas pressures (i.e. fewer HHG emitters) [55].

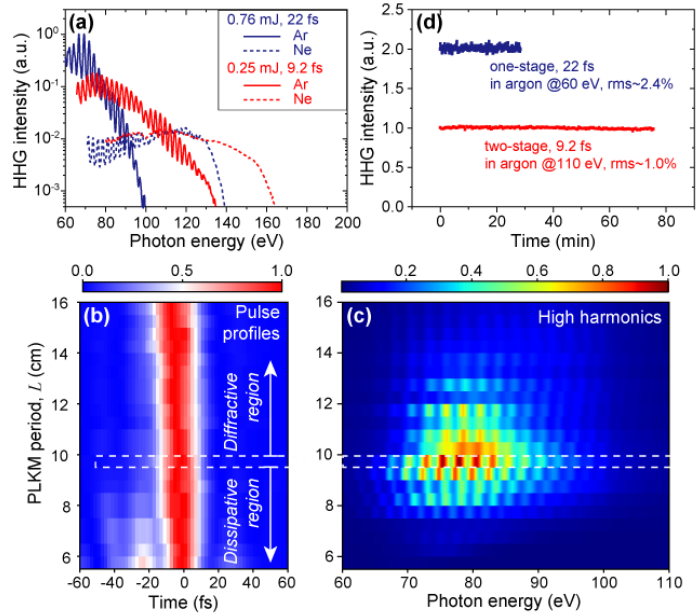

Fig. 5. (a) The HHG spectra driven by compressed pulses in argon and neon. (b) The pulse intensity profiles measured by SHG-FROG under different PLKM periods *L* of the second-stage compressor. The optimum condition is determined to be *L*=9.5 cm. The dissipative and diffractive regions are labelled. (c) The HHG spectra excited by the pulses in (d) under different *L*. (d) The long-term stability of a single harmonic order in the cut-off region of HHG driven by the compressed pulses in argon.

In Fig. 5(b) and (c), we present the temporal profiles of the few-cycle pulses and the corresponding harmonic spectra in argon, respectively, for different *L* of the second-stage PLKM compressor. The pulse energy remains constant (~0.25 mJ). When *L* is short (5 - 8 cm, the "*dissipative region*"), the complex space-time coupling gives rise to higher-order dispersion, which cannot be appropriately compensated. As a result, multiple pedestals can be observed in the pulse temporal shapes [Fig. 5(b)], with the HHG brightness and cut-off energy both reduced accordingly [Fig. 5(c)]. On the other hand, in the "*diffractive region*" where *L* is long (>11 cm), the nonlinear phase and SCG bandwidth reduces, leading to the long pulse durations (e.g. $\tau \approx 14$ fs at *L*=16 cm) and weak HHG emission [Fig. 5(c)]. Furthermore, complex spatio-spectral shapes of the HHG radiation can be observed in the non-soliton regions (see Supplementary Section S6). These results clearly demonstrate that the resultant condition determined with our recipe (*l*=100 μm and *L*=9.5 cm) can indeed give rise to the optimum few-cycle-pulse compression in both space and time.

The measurements on the long-term stability of the HHG radiation driven by the single-stage and the two-stage compressors are present in Fig. 5(d). We note that no pointing-stabilization apparatus was implemented in either case. The flux of an individual harmonic order generated in argon in the cut-off region (~60 eV for 22 fs and ~110 eV for 9.2 fs) is inspected in each case. The measurement yields an rms deviation of as low as 1% over a time period more than 1 hour. This result highlights the exceptional robustness and stability of our method. We believe it can be useful in physics, chemistry and biology laboratories, in which coherent and ultrashort X-ray sources with long-term stability are in demand.

## 3. Conclusion

In summary, ~9.2 fs few-cycle pulses are generated with high efficiency (>50%) and high spatio-spectral homogeneity (~0.87) by compressing the 1030 nm 170 fs pulses from a Yb:KGW laser using an all-solid-state free-space pulse compressor. This is achieved by managing the nonlinear beam propagation in PLKM into soliton modes and confining the nonlinear phase on each material layer to be less than 1.0 rad. We develop a complete recipe for constructing and optimizing the PLKM compressors, and demonstrate that it can be flexibly implemented for a wide range of laser conditions of the Yb lasers. The compressor exhibits high efficiency, high spatiotemporal quality and exceptional stability. In combination with the power and repetition-rate scalable Yb lasers, we believe it will enable new applications in ultrafast science and extreme nonlinear optics.

**Funding.** National Natural Science Foundation of China (1874121); the Shanghai Municipal Science and Technology Basic Research Project (19JC1410900). M.M. and H.K. gratefully acknowledge funding from a Physics Frontier Center (PHY 1734006) for the experimental techniques, and an AFOSR MURI (FA9550-16-1-0121) for theory.

**Acknowledgments.** We thank Andrius Baltuska, Seth Cousin, Michaël Hemmer and Daniel Carlson for helpful discussions.

## References

1. A. Kaldun, S. Donsa, H. Wei, R. Pazourek, S. Nagele, C. Ott, C. D. Lin, and T. Pfeifer, "Observing the ultrafast buildup of a Fano resonance in the time domain," Science **166**, 162–166 (2016).

2. M. Ossiander, F. Siegrist, V. Shirvanyan, R. Pazourek, A. Sommer, T. Latka, A. Guggenmos, S. Nagele, J. Feist, J. Burgdörfer, R. Kienberger, and M. Schultze, "Attosecond correlation dynamics," Nat. Phys. **13**, 280–285 (2017).
3. E. Goulielmakis, M. Schultze, M. Hofstetter, V. S. Yakovlev, J. Gagnon, M. Uiberacker, A. L. Aquila, E. M. Gullikson, D. T. Attwood, and R. Kienberger, "Single-Cycle Nonlinear Optics," Science **320**, 1614–1617 (2008).
4. F. Calegari, D. Ayuso, A. Trabattoni, L. Belshaw, S. De Camillis, S. Anumula, F. Frassetto, L. Poletto, A. Palacios, P. Decleva, J. B. Greenwood, F. Martín, and M. Nisoli, "Ultrafast electron dynamics in phenylalanine initiated by attosecond pulses," Science **346**, 336–339 (2014).
5. X. Shi, C. T. Liao, Z. Tao, E. Cating-Subramanian, M. M. Murnane, C. Hernández-García, and H. C. Kapteyn, "Attosecond light science and its application for probing quantum materials," J. Phys. B At. Mol. Opt. Phys. **53**, 184008 (2020).
6. A. Rudenko, K. Zrost, B. Feuerstein, V. L. B. De Jesus, C. D. Schröter, R. Moshammer, and J. Ullrich, "Correlated multielectron dynamics in ultrafast laser pulse interactions with atoms," Phys. Rev. Lett. **93**, 253001 (2004).
7. A. Schiffrin, T. Paasch-Colberg, N. Karpowicz, V. Apalkov, D. Gerster, S. Mühlbrandt, M. Korbman, J. Reichert, M. Schultze, S. Holzner, J. V Barth, R. Kienberger, R. Ernstorfer, V. S. Yakovlev, M. I. Stockman, and F. Krausz, "Optical-field-induced current in dielectrics," Nature **493**, 70–74 (2013).
8. A. L. Cavalieri, N. Müller, T. Uphues, V. S. Yakovlev, A. Baltuška, B. Horvath, B. Schmidt, L. Blümel, R. Holzwarth, S. Hendel, M. Drescher, U. Kleineberg, P. M. Echenique, R. Kienberger, F. Krausz, and U. Heinzmann, "Attosecond spectroscopy in condensed matter.," Nature **449**, 1029–1032 (2007).
9. M. Schultze, K. Ramasesha, C. D. Pemmaraju, S. A. Sato, D. Whitmore, A. Gandman, J. S. Prell, L. J. Borja, D. Prendergast, K. Yabana, D. M. Neumark, and S. R. Leone, "Attosecond band-gap dynamics in silicon," Science **346**, 1348–1352 (2014).
10. F. Siegrist, J. A. Gessner, M. Ossiander, C. Denker, Y. P. Chang, M. C. Schröder, A. Guggenmos, Y. Cui, J. Walowski, U. Martens, J. K. Dewhurst, U. Kleineberg, M. Münzenberg, S. Sharma, and M. Schultze, "Light-wave dynamic control of magnetism," Nature **571**, 240–244 (2019).
11. P. B. Corkum, "Plasma perspective on strong-field multiphoton ionization," Phys. Rev. Lett. **71**, 1994–1997 (1993).
12. T. Popmintchev, M. C. Chen, D. Popmintchev, P. Arpin, S. Brown, S. Ališauskas, G. Andriukaitis, T. Balčiunas, O. D. Mücke, A. Pugzlys, A. Baltuška, B. Shim, S. E. Schrauth, A. Gaeta, C. Hernández-García, L. Plaja, A. Becker, A. Jaron-Becker, M. M. Murnane, and H. C. Kapteyn, "Bright coherent ultrahigh harmonics in the kev x-ray regime from mid-infrared femtosecond lasers," Science **336**, 1287–1291 (2012).
13. T. Popmintchev, M.-C. Chen, P. Arpin, M. M. Murnane, and H. C. Kapteyn, "The attosecond nonlinear optics of bright coherent X-ray generation," Nat. Photonics **4**, 822–832 (2010).
14. D. Guénot, D. Gustas, A. Vernier, B. Beaurepaire, F. Böhle, M. Bocoum, M. Lozano, A. Jullien, R. Lopez-Martens, A. Lifschitz, and J. Faure, "Relativistic electron beams driven by kHz single-cycle light pulses," Nat. Photonics **11**, 293–296 (2017).
15. W. H. Knox, R. L. Fork, M. C. Downer, R. H. Stolen, C. V. Shank, and J. A. Valdmanis, "Optical pulse compression to 8 fs at a 5-kHz repetition rate," Appl. Phys. Lett. **46**, 1120–1121 (1985).
16. A. Dubietis, R. Butkus, and A. P. Piskarskas, "Trends in chirped pulse optical parametric amplification," IEEE J. Sel. Top. Quantum Electron. **12**, 163–172 (2006).
17. C. V. Shank, R. L. Fork, R. Yen, R. H. Stolen, and W. J. Tomlinson, "Compression of femtosecond optical pulses," Appl. Phys. Lett. **40**, 761–763 (1982).
18. B. Nikolaus and D. Grischkowsky, "12× Pulse Compression Using Optical Fibers," Appl. Phys. Lett. **42**, 1–2 (1983).
19. M. Nisoli, S. De Silvestri, and O. Svelto, "Generation of high energy 10 fs pulses by a new pulse compression technique," Appl. Phys. Lett. **68**, 2793–2795 (1996).
20. M. Nisoli, S. De Silvestri, O. Svelto, R. Szipöcs, K. Ferencz, C. Spielmann, S. Sartania, and F. Krausz, "Compression of high-energy laser pulses below 5 fs," Opt. Lett. **22**, 522 (1997).
21. N. Milosevic, G. Tempea, and T. Brabec, "Optical pulse compression: bulk media versus hollow waveguides," Opt. Lett. **25**, 672–674 (2000).
22. C. H. Lu, T. Witting, A. Husakou, M. Vrakking, A. H. Kung, and F. Furch, "Sub-4 fs laser pulses at high average power and high repetition rate from an all-solid-state setup," Opt. Express **26**, 1267–1269 (2018).
23. M. Seidel, G. Arisholm, J. Brons, V. Pervak, and O. Pronin, "All solid-state spectral broadening: an average and peak power scalable method for compression of ultrashort pulses," Opt. Express **24**, 9412 (2016).
24. O. Pronin, M. Seidel, F. Lücking, J. Brons, E. Fedulova, M. Trubetskov, V. Pervak, A. Apolonski, T. Udem, and F. Krausz, "High-power multi-megahertz source of waveform-stabilized few-cycle light," Nat. Commun. 6998 (2015).
25. T. Eidam, S. Hanf, T. V. Andersen, E. Seise, C. Wirth, T. Schreiber, T. Gabler, J. Limpert, and A. Tünnermann, "830 W average power femtosecond fiber CPA system," Opt. Lett. **35**, 94–96 (2010).
26. J.-P. Negel, A. Voss, M. A. Ahmed, D. Bauer, D. Sutter, A. Killi, and T. Graf, "11 kW average output power from a thin-disk multipass amplifier for ultrashort laser pulses," Opt. Lett. **38**, 5442 (2013).

27. P. Russbueldt, T. Mans, J. Weitenberg, H. D. Hoffmann, and R. Poprawe, "Compact diode-pumped 11 kW Yb:YAG Innoslab femtosecond amplifier," Opt. Lett. **35**, 4169 (2010).
28. J. Schulte, T. Sartorius, J. Weitenberg, A. Vernaleken, and P. Russbueldt, "Nonlinear pulse compression in a gas-filled multipass cell," Opt. Lett. **41**, 4511 (2016).
29. T. Nagy, S. Hädrich, P. Simon, A. Blumenstein, N. Walther, R. Klas, J. Buldt, H. Stark, S. Breitkopf, P. Jójárt, I. Seres, Z. Várallyay, T. Eidam, and J. Limpert, "Generation of three-cycle multi-millijoule laser pulses at 318 W average power," Optica **6**, 1423–1424 (2019).
30. R. Klas, W. Eschen, A. Kirsche, J. Rothhardt, and J. Limpert, "Generation of coherent broadband high photon flux continua in the XUV with a sub-two-cycle fiber laser," Opt. Express **28**, 6188 (2020).
31. L. Lavenu, M. Natile, F. Guichard, Y. Zaouter, X. Delen, M. Hanna, E. Mottay, and P. Georges, "Nonlinear pulse compression based on a gas-filled multipass cell," Opt. Lett. **43**, 2252 (2018).
32. F. Köttig, D. Schade, J. R. Koehler, P. S. J. Russell, and F. Tani, "Efficient single-cycle pulse compression of an ytterbium fiber laser at 10 MHz repetition rate," Opt. Express **28**, 9099–9110 (2020).
33. Y. G. Jeong, R. Piccoli, D. Ferachou, V. Cardin, M. Chini, S. Hädrich, J. Limpert, R. Morandotti, F. Légaré, B. E. Schmidt, and L. Razzari, "Direct compression of 170-fs 50-cycle pulses down to 1.5 cycles with 70% transmission," Sci. Rep. **8**, 11794 (2018).
34. S. Hädrich, M. Kienel, M. Müller, A. Klenke, J. Rothhardt, R. Klas, T. Gottschall, T. Eidam, A. Drozdy, P. Jójárt, Z. Várallyay, E. Cormier, K. Osvay, A. Tünnermann, and J. Limpert, "Energetic sub-2-cycle laser with 216 W average power," Opt. Lett. **41**, 4332 (2016).
35. C. Grebing, M. Müller, J. Buldt, H. Stark, and J. Limpert, "Kilowatt-average-power compression of millijoule pulses in a gas-filled multi-pass cell," Opt. Lett. **45**, 6250 (2020).
36. G. Fan, P. A. Carpeggiani, Z. Tao, G. Coccia, R. Safaei, E. Kaksis, A. Pugzlys, F. Légaré, B. E. Schmidt, and A. Baltuška, "70 mJ nonlinear compression and scaling route for an Yb amplifier using large-core hollow fibers," Opt. Lett. **46**, 896 (2021).
37. F. Emaury, C. J. Saraceno, B. Debord, D. Ghosh, A. Diebold, F. Gèrôme, T. Südmeyer, F. Benabid, and U. Keller, "Efficient spectral broadening in the 100-W average power regime using gas-filled kagome HC-PCF and pulse compression," Opt. Lett. **39**, 6843 (2014).
38. P. Balla, A. Bin Wahid, I. Sytcevich, C. Guo, A. L. Viotti, L. Silletti, A. Cartella, S. Alisauskas, H. Tavakol, U. Grosse-Wortmann, A. Schönberg, M. Seidel, A. Trabattoni, B. Manschwetus, T. Lang, F. Calegari, A. Couairon, A. L'Huillier, C. L. Arnold, I. Hartl, and C. M. Heyl, "Post-compression of picosecond pulses into the few-cycle regime," Opt. Lett. **45**, 2572–2575 (2020).
39. J. E. Beetar, S. Gholam-Mirzaei, and M. Chini, "Spectral broadening and pulse compression of a 400 µJ, 20 W Yb:KGW laser using a multi-plate medium," Appl. Phys. Lett. **112**, 051102 (2018).
40. K. Fritsch, M. Poetzlberger, V. Pervak, J. Brons, and O. Pronin, "All-solid-state multipass spectral broadening to sub-20 fs," Opt. Lett. **43**, 4643 (2018).
41. J. Weitenberg, A. Vernaleken, J. Schulte, A. Ozawa, T. Sartorius, V. Pervak, H.-D. Hoffmann, T. Udem, P. Russbüldt, and T. W. Hänsch, "Multi-pass-cell-based nonlinear pulse compression to 115 fs at 75 µJ pulse energy and 300 W average power," Opt. Express **25**, 20502 (2017).
42. C.-H. Lu, W.-H. Wu, S.-H. Kuo, J.-Y. Guo, M.-C. Chen, S.-D. Yang, and A. H. Kung, "Greater than 50 times compression of 1030 nm Yb:KGW laser pulses to single-cycle duration," Opt. Express **27**, 15638 (2019).
43. N. Ishii, P. Xia, T. Kanai, and J. Itatani, "Optical parametric amplification of carrier-envelope phase-stabilized mid-infrared pulses generated by intra-pulse difference frequency generation," Opt. Express **27**, 11447 (2019).
44. S. Zhang, Z. Fu, B. Zhu, G. Fan, Y. Chen, S. Wang, Y. Liu, A. Baltuska, C. Jin, C. Tian, and Z. Tao, "Solitary beam propagation in periodic layered Kerr media enables high-efficiency pulse compression and mode self-cleaning," Light Sci. Appl. **10**, 53 (2021).
45. C.-H. Lu, Y.-J. Tsou, H.-Y. Chen, B.-H. Chen, Y.-C. Cheng, S.-D. Yang, M.-C. Chen, C.-C. Hsu, and A. H. Kung, "Generation of intense supercontinuum in condensed media," Optica **1**, 400–406 (2014).
46. P. He, Y. Liu, K. Zhao, H. Teng, X. He, P. Huang, H. Huang, S. Zhong, Y. Jiang, S. Fang, X. Hou, and Z. Wei, "High-efficiency supercontinuum generation in solid thin plates at 01 TW level," Opt. Lett. **42**, 474 (2017).
47. L. Bergé, S. Skupin, R. Nuter, J. Kasparian, and J. P. Wolf, "Ultrashort filaments of light in weakly ionized, optically transparent media," Reports Prog. Phys. **70**, 1633–1713 (2007).
48. A. M. Weiner, R. H. Stolen, and J. P. Heritage, "Self-phase modulation and optical pulse compression influenced by stimulated Raman scattering in fibers," J. Opt. Soc. Am. B **5**, 364 (1988).
49. S. C. Pinault and M. J. Potasek, "Frequency broadening by self-phase modulation in optical fibers," J. Opt. Soc. Am. B **2**, 1318–1319 (1985).
50. R. Trebino, K. W. DeLong, D. N. Fittinghoff, J. N. Sweetser, M. A. Krumbügel, B. A. Richman, and D. J. Kane, "Measuring ultrashort laser pulses in the time-frequency domain using frequency-resolved optical gating," Rev. Sci. Instrum. **68**, 3277–3295 (1997).
51. A. A. Zozulya, S. A. Diddams, A. G. Van Engen, and T. S. Clement, "Propagation dynamics of intense femtosecond pulses: Multiple splittings, coalescence, and continuum generation," Phys. Rev. Lett. **82**, 1430–1433 (1999).

52. I. P. Christov, J. Zhou, J. Peatross, A. Rundquist, M. M. Murnane, and H. C. Kapteyn, "Nonadiabatic effects in high-harmonic generation with ultrashort pulses," Phys. Rev. Lett. **77**, 1743–1746 (1996).
53. G. Tempea, M. Geissler, M. Schnürer, and T. Brabec, "Self-Phase-Matched High Harmonic Generation," Phys. Rev. Lett. **84**, 4329–4332 (2000).
54. C. Ding, W. Xiong, T. Fan, D. D. Hickstein, X. Zhang, M. Walls, M. M. Murnane, and H. C. Kapteyn, "High flux coherent super-continuum soft X-ray source driven by a single-stage , 10mJ , Ti : sapphire amplifier-pumped OPA," Opt. Express **22**, 6194–6202 (2014).
55. P. Arpin, T. Popmintchev, N. L. Wagner, A. L. Lytle, O. Cohen, H. C. Kapteyn, and M. M. Murnane, "Enhanced high harmonic generation from multiply ionized argon above 500eV through laser pulse self-compression," Phys. Rev. Lett. **103**, 143901 (2009).